# It's Not the AI – It's Each of Us!
# Ten Commandments for the Wise & Responsible Use of AI

Barbara Steffen[1], Edward A. Lee[2], Moshe Y. Vardi[3], and Bernhard Steffen[4]

[1]**METAFrame Technologies GmbH, Dortmund, Germany**
barbara.steffen@metaframe.de

[2]**UC Berkeley, Berkeley, USA**
eal@berkeley.edu

[3]**Rice University, Houston, USA**
vardi@rice.edu

[4]**TU Dortmund University, Dortmund, Germany**
steffen@cs.tu-dortmund.de

Artificial intelligence (AI) is no longer futuristic; it is a daily companion shaping our private and work lives. While AI simplifies our lives, its rise also invites us to rethink who we are – and who we wish to remain – as humans. Even if AI does not think, feel, or desire, it learns from our behavior, mirroring our collective values, biases, and aspirations. The question, then, is not what AI is, but what we are allowing it to become through data, computing power, and other parameters "teaching" it – and, even more importantly, who we are becoming through our relationship with AI.

As the EU AI Act [5,6] and the Vienna Manifesto on Digital Humanism [22,23] emphasize, technology must serve human dignity, social well-being, and democratic accountability. In our opinion, responsible use of AI is not only a matter of code nor law, but also of conscientious practice: how each of us engages and teaches others to use AI at home and at work. The here proposed Ten Commandments for the Wise and Responsible Use of AI are meant as guideline for this very engagement. They closely align with Floridi and Cowls' [7] five guiding principles for AI in society – beneficence, non-maleficence, autonomy, justice, and explicability.

## It's on Individuals to Navigate AI's Double-Edged Sword

AI addresses – and amplifies – human weaknesses. Considering human's bounded rationality (Simon, [17]), today's pervasive information overload, and rapid change, it is not surprising that we prefer ease, chase simple solutions, delegate to technology, and resist uncertainty. These dispositions make AI compelling – and leave us vulnerable to the very systems we create. We become both its eager adopters and unintended casualties.

The love of convenience drives people to embrace AI's efficiency and instant accessibility. In an age that rewards productivity and speed, that affection for convenience becomes a trap: we let systems think for us, mistaking comfort for competence. Carr [2] warns that this convenience erodes our capacity for deep reflection, a concern echoed by Steffen et al. [20], who caution that delegating cognitive effort to AI gradually undermines human autonomy and intellectual resilience. Comfort becomes harmful when it clouds our willingness to think or to engage in slow, laborious thought processes.

Kahneman [8] demonstrates that people often prefer a *certain negative outcome* to facing uncertainty, revealing a deep-seated discomfort with the unknown. This tendency – to favor a bad certainty over an uncertain possibility – is well established in behavioral economics and psychology, notably through research on loss aversion (Kahneman & Tversky, [9]), ambiguity aversion (Ellsberg, [4]), and status quo bias (Samuelson & Zeckhauser, [16]).

AI caters to this weakness by delivering confident answers, soothing our anxiety even when its outputs are merely probabilistic guesses. The interviews in [19] highlight the risk that this craving for certainty promotes blind trust in AI tools, replacing critical dialogue and human oversight. Early evidence shows lower brain engagement when people write with ChatGPT versus unaided work (Kosmyna et al., [12]). At work, this fuels "workslop" – polished but low-value output that drags productivity (Niederhoffer et al., [13]). Even worse, as O'Neil [14] warns, for non-experts it becomes nearly impossible to distinguish what sounds right from what is right.

Together, these two tendencies form what Steffen [21] calls the "boiling frog" syndrome, a metaphor for the gradual, almost imperceptible loss of critical awareness. Like a frog placed in cold water that is slowly heated, we fail to notice the creeping danger until it is too late. The immediate threat is not a sudden technological overthrow but a gradual risk of increasing complacency; convenience and certainty can quietly erode curiosity, discernment, and moral agency. It is up to each of us to stay alert, cultivate skill, and educate ourselves to use AI consciously and responsibly.

## The Individual's Responsibilities in Private Life

Children, elderly, and other vulnerable groups are facing technologies and dangers they are not prepared for. It is our responsibility – as parents, children, and friends – to lay the foundation (see Figure 1). Take children, for example: they must learn to master the basics – speaking, reading, writing, thinking, reflecting, debating – despite the temptation to outsource babysitting to, e.g., social media, TV, and games. In fact, today, children encounter AI daily – in voice assistants, recommendation systems, entertainment offers, and educational tools – often well before they had the chance to build core competences and self-protection guards.

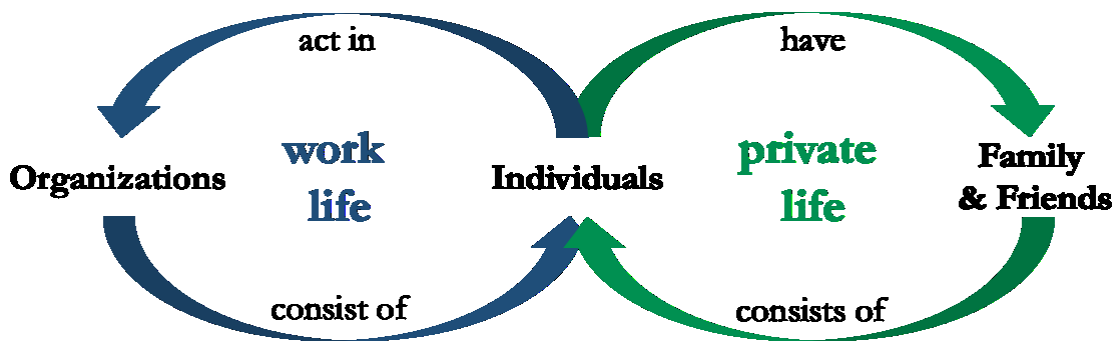

**Figure 1. Interplay and Reinforcement Loop of an Individual's Areas of Responsibility** (adapted from Steffen, 2024).

Kasneci et al. [10] show that AI can enhance learning when used to stimulate inquiry rather than deliver fixed answers. Without guidance, however, children and vulnerable groups risk becoming passive consumers of algorithmic output. Therefore, it is important that parents guide their children to adopt AI in a balanced, reflected way.

More generally, each of us must cultivate reflective AI use: talk through when to trust a digital source, how bias creeps in, and how to navigate tools built on statistical prediction. Such conversations build metacognition – the capacity to reflect on one's own thinking. Parents, schools, colleges, and adult-education programs can reinforce this by teaching not only technical skills but also media, moral, and civic literacy. As Luckin et al. [11] argue, the aim in the AI era is to cultivate *digital wisdom*: using AI to extend curiosity, not replace it. The development of digital wisdom begins at playgrounds, kitchen tables, and over a beer, where children and non-experts learn to balance trust with healthy skepticism.

## The Individual's Responsibilities in Work Life

While families and friends shape habits of mind, organizations build the structures in which those habits operate. The EU AI Act calls for human-centric, trustworthy AI – ethics woven into innovation, responsible adoption, and investment in AI literacy – grounded in transparency and accountability. Also in the workplace, conscious AI adoption is essential (see Figure 1). In the capacity as an organization's AI lead, boss, early adopter, or expert, it is on you to help colleagues understand what AI is, which tools are safe, what uses are permitted, and how to use them well. And if you are none of the above, it is on you to ask for AI training.

The current AI hype implies that "more AI is always better", yet failed pilots and rising "workslop" show that AI is a tool with strengths and limits. Conscious use is therefore both an economic necessity and a prerequisite for regulatory compliance, safety, and ethics. Moving beyond compliance means turning responsibility into capability: establish interdisciplinary AI governance boards, audit systems for bias and impact, and reward ethical awareness. In education and health care, responsible practice protects not only personal data but dignity; in business, it builds trust and long-term resilience. Ultimately, organizational maturity in AI ethics mirrors human maturity; it grows through reflection, humility, and shared values.

## Ten Commandments for the Wise and Responsible Use of AI

The following ten commandments guide each of us in our relationship with – and adoption of – AI. They emphasize the responsibility of each of us for our actions and our duties as members of society, especially within our private and work roles: At the end, organizations and society are driven by the actions of individuals (see Figure 1). In particular, the ten commandments clarify what it means to *protect humanity (why)*, *ensure non-negotiable safeguards (what)*, *act purposefully (how),* and *continuously evaluate the Purpose–Impact fit (reflect)* (see Figure 2). Taken together, they are a step towards operationalizing the vision of the EU AI Act by embedding human-centered values into everyday practice.

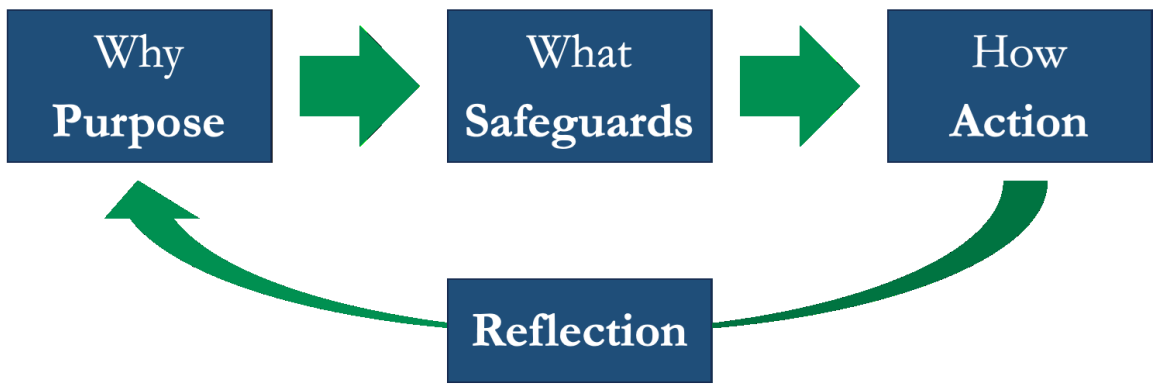

**Figure 2. The Continuous Cycle of the Wise and Responsible Use of AI** (adapted from Steffen, 2024).

## Purpose: Protect Humanity

1. **Foster & Protect Human Capabilities**
   Keep curiosity, critical thinking, empathy, and creativity at the center. Use AI to learn and reflect – not to outsource your mind.
2. **Limit AI Dependency**
   Regularly assess where reliance is forming and the risks it creates (now and later). Set limits and consider alternatives to prevent falling victim to the lock-in effect.
3. **Consider *all* the costs**
   Choose AI use cases, models, and workflows that minimize environmental and social costs across their lifecycle.

## Safeguards: Ensure Non-Negotiable Principles

4. **Promote Fairness & Inclusivity**
   Detect and reduce bias so that outcomes serve everyone – especially those historically excluded – and establish availability and benefit of AI throughout society.
5. **Protect Privacy & Data Security**
   Protect personal data – with consent, confidentiality, and special care for vulnerable individuals and groups.
6. **Make AI Use Visible & Discussable**
   Disclose when, where, and how AI is used. Transparency builds trust and enables conscious use and critique.

## Action: Proceed Purposefully

7. **Adopt AI Consciously & Deliberately**
   Ask: Why use it? What does it enable or replace? Does it truly serve human interests?
8. **Clarify Accountability Before Action**
   Keep responsibility human. Treat AI outputs as proposals – verify reasoning, context, and ethics before implementation.
9. **Prioritize Values over Efficiency**
   Seek understanding, meaning, and insight first. Speed is helpful – but not at the cost of quality or human values.

**Reflection: Continuously Evaluate the Purpose–Impact Fit**

10. **Commit to Oversight: Measure, Learn, and Control**
    Monitor impacts and adjust continuously. Evolve practices to keep benefits high and harms low – keep the "frog" alive.

# Conclusion

AI mirrors the society that builds it. It can illuminate human creativity or amplify complacency. The Ten Commandments for the Wise and Responsible Use of AI offer a framework to guide our responsibility – for ourselves, in our private and work lives. They are grounded in our observations over recent years, including two iterative feedback cycles with interdisciplinary experts at AISoLA 2025; Digital wisdom requires moral awareness and conscientious practice — the courage to engage critically with the systems we create, and the care to teach and model their responsible use in everyday life.

The commandments gain strength when understood as interdependent; each dimension reinforces the others (see Figure 1). Reflective homes shape ethical employees; transparent organizations empower people who, in turn, build trust in technology in private life. They are in line with the EU AI Act which aims to scale this interdependence by fostering ecosystems of trust where innovation and ethics coevolve. Responsible AI is not a technical endpoint but a cultural practice requiring us – citizens, educators, leaders, and policymakers – to keep progress human-centered.

Existential risks – from super intelligent AI escaping human control to deliberate misuse – remain serious concerns. Scholars such as Russell [15] and Bostrom [1] have explored scenarios in which an uncontrollable superintelligence could emerge, with potentially irreversible consequences. As Bostrom cautions, "Once unfriendly superintelligence exists, it would prevent us from replacing it or changing its preferences. Our fate would be sealed." This fear is echoed by Yudkowsky and Soares [24], who starkly warn, "If anyone builds it, everyone dies." Reflecting such growing anxiety, the Center for AI Safety [3] has urged that "mitigating the risk of extinction from AI should be a global priority alongside other societal-scale risks such as pandemics and nuclear war."

While the probability of these scenarios remains uncertain, they demand vigilant attention. Whatever one's estimate of the risks, the Ten Commandments insist that consciousness, responsibility, transparency, openness, collaboration, and human oversight are non-negotiable safeguards against both present and future dangers.

To "keep the frog alive", we will revisit the commandments regularly to reflect on their relevance, timeliness, and completeness. Ultimately, it's not up to the AI; it's up to each of us – our conscious, established, and lived use of AI.

## References

1. Bostrom, N., *Superintelligence: Paths, Dangers, Strategies*. Oxford University Press, 2014

2. Carr, N.. *The Shallows: What the Internet Is Doing to Our Brains*. W. W. Norton, 2010

3. Center for AI Safety, 2023, *Statement on AI Risk*. https://www.safe.ai/statement-on-ai-risk

4. Ellsberg, D.. *Risk, Ambiguity, and the Savage Axioms*. Quarterly Journal of Economics, 75(4), 643–669, 1961

5. EU-Artificial Intelligence Act, https://artificialintelligenceact.eu

6. EU-Artificial Intelligence Act: First Regulation on Artificial Intelligence, https://www.europarl.europa.eu/topics/en/article/20230601STO93804/eu-ai-act-first-regulation-on-artificial-intelligence

7. Floridi, L., & Cowls, J. *A Unified Framework of Five Principles for AI in Society*. Harvard Data Science Review, 2019

8. Kahneman, D., *Thinking, Fast and Slow*. Farrar, Straus, and Giroux, 2011

9. Kahneman, D., & Tversky, A.. *Prospect Theory: An Analysis of Decision under Risk*. Econometrica, 47(2), 263–291, 1979

10. Kasneci, E., Sessler, K., et al., 2023, *ChatGPT for Good? On Opportunities and Challenges of Large Language Models for Education*. Learning and Individual Differences, 103, 102274. https://doi.org/10.1016/j.lindif.2023.102274

11. Luckin, R., George, K., Cukurova, M., 2022, *AI for School Teachers*. CRC Press https://doi.org/10.1201/9781003193173

12. Kosmyna, N., Hauptmann, E., Yuan, Y.T., Situ, J., Liao, X.H., Beresnitzky, A.V., Braunstein, I. and Maes, P.. *Your Brain on ChatGPT: Accumulation of Cognitive Debt when using an AI Assistant for Essay Writing Task*. arXiv preprint arXiv:2506.08872, 2025

13. Niederhoffer, K. et al., *AI-Generated 'Workslop' Is Destroying Productivity*. Harvard Business Review, 2025

14. O'Neil, C., *Weapons of Math Destruction: How Big Data Increases Inequality and Threatens Democracy*, Crown Publishing Group (NY), 2016

15. Russell, S., *Human Compatible: Artificial Intelligence and the Problem of Control*, Penguin Publishing Group, 2020

16. Samuelson, W., & Zeckhauser, R. (1988). *Status Quo Bias in Decision Making*. Journal of Risk and Uncertainty, 1(1), 7–59, 1988

17. Simon, H. A., *"Bounded rationality," Utility and probability*. Springer, pp. 15–18, 1990

18. Steffen, B., *Alignment-Driven Adaptation Process & Tool* (ADAPT): Towards continuous and holistic adaptation of organizations (Doctoral dissertation, Dissertation, Potsdam, Universität Potsdam), 2024

19. Steffen, B., Lee, E. A., & Steffen, B. (Eds.), *Let's Talk AI: Interdisciplinarity Is a Must.* Springer Nature, 2025

20. Steffen, B., Lee, A. E., Steffen, B. (Eds.), "*Let's Talk AI: Impressions and Thoughts After 30 Interviews.*" In Steffen et al., 2025

21. Steffen, B., *How Hot is the Water?* In Steffen et al., 2025.

22. Vienna Manifesto on Digital Humanism, 2019, TU Wien Informatics, https://dighum.ec.tuwien.ac.at/vienna-manifesto/

23. Werthner, H., Prem, E., Lee, A. E., Ghezzi, C., *Perspectives on Digital Humanism*. Springer International Publishing, 2022

24. Yudkowsky, E., Soares, N., *If Anyone Builds It, Everyone Dies: Why Superhuman AI Would Kill Us All*. Little, Brown and Company 2025